\documentclass[reprint,amsmath,amssymb,pra,aps,nofootinbib,superscriptaddress,floatfix]{revtex4-2}
\pdfoutput=1
\usepackage[dvipsnames]{xcolor}
\definecolor{dcyan}{RGB}{0,100,100}
\usepackage[T1]{fontenc}
\usepackage[utf8]{inputenc}
\usepackage{amsmath, amssymb}
\usepackage{bbold}
\usepackage{bm}
\usepackage{tipa}
\usepackage{subfigure}
\usepackage{blindtext}
\usepackage{graphicx}
\usepackage{tabularx}
\usepackage{textgreek}
\usepackage{xfrac}
\usepackage{units}
\usepackage{comment}
\usepackage{float}
\usepackage{placeins}

\usepackage[sectionbib]{bibunits}
\defaultbibliographystyle{naturemag}
\defaultbibliography{FinalBib}

\usepackage{hyperref}
\usepackage{xcolor}
\definecolor{green_cust}{RGB}{0,154,85}
\definecolor{red_cust}{RGB}{173,49,54}
\definecolor{blue_cust}{RGB}{0,103,148}
\hypersetup{colorlinks=true, linkcolor=red_cust, citecolor=green_cust, filecolor=blue_cust, urlcolor=blue_cust}

\makeatletter 

\renewcommand\onecolumngrid{
\do@columngrid{one}{\@ne}%
\def\set@footnotewidth{\onecolumngrid}
\def\footnoterule{\kern-6pt\hrule width 1.5in\kern6pt}%
}

\renewcommand\twocolumngrid{
        \def\footnoterule{
        \dimen@\skip\footins\divide\dimen@\thr@@
        \kern-\dimen@\hrule width.5in\kern\dimen@}
        \do@columngrid{mlt}{\tw@}
}%

\makeatother    

\usepackage{soul}

\newcommand{\Figref}[1]{Fig.~\hyperref[#1]{\ref{#1}}}

\graphicspath{{./figures/}}

\begin{document}
\title{A Mega-FPS low light camera}

\author{Bowen Li}
\affiliation{Department of Physics, Stanford University, Stanford, CA}

\author{Lukas Palm}
\affiliation{Department of Physics, The University of Chicago and the James Franck Institute, Chicago, IL}

\author{Marius J\"urgensen}
\affiliation{Department of Physics, Stanford University, Stanford, CA}

\author{Yiming Cady Feng}
\affiliation{Department of Applied Physics, Stanford University, Stanford, CA}

\author{Markus Greiner}
\affiliation{Department of Physics, Harvard University, Cambridge, MA}

\author{Jon Simon}
\affiliation{Department of Physics, Stanford University, Stanford, CA}
\affiliation{Department of Applied Physics, Stanford University, Stanford, CA}

\date{\today}
\begin{bibunit}

\begin{abstract}
From biology and astronomy to quantum optics, there is a critical need for high frame rate, high quantum efficiency imaging. In practice, most cameras only satisfy one of these requirements. Here we introduce interlaced fast kinetics imaging, a technique that allows burst video acquisition at frame rates up to 3.33~Mfps using a commercial EMCCD camera with single-photon sensitivity. This approach leverages EMCCD's intrinsic fast row transfer dynamics by introducing a tilted lens array into the imaging path, creating a spatially distributed grid of exposed pixels, each aligned to its own column of the sensor. The remaining unexposed pixels serve as in-situ storage registers, allowing subsequent frames to be captured after just one row shift operation. Our interlaced fast kinetics camera maintains 50\% contrast for square wave intensity modulation frequencies up to 1.61~MHz. We provide benchmarks of the video performance by capturing two dimensional videos of spatially evolving patterns that repeat every~2$\mu$s, with spatial resolution of 11$\times$15 pixels. Our approach is compatible with commercial EMCCDs and opens a new route to ultra-fast imaging at single-photon sensitivity with applications from fast fluorescence imaging to photon correlation measurement.

\end{abstract}
\maketitle

\section{Introduction}
\label{sec:intro}
Fast single-photon imaging is rapidly becoming a pivotal technology in numerous scientific disciplines, from mid-circuit readout in quantum computing~\cite{singh_mid-circuit_2023, bluvstein_logical_2024}, to wide-field fluorescence lifetime imaging microscopy~\cite{suhling_fluorescence_2015, hirvonen_fast_2020} in biology, speckle interferometry and lucky imaging in astronomy~\cite{smith_investigation_2009, strakhov_speckle_2023}, and quantum-enhanced sensing~\cite{moreau_imaging_2019, defienne_advances_2024}. The silicon detector arrays employed in visible-light cameras typically require a choice between high speed and single-photon sensitivity: On one hand, digital cameras with frame rates up to $10^8$~fps have been developed~\cite{mochizuki_64_2015,etoh_light--flight_2019,suzuki_over_2020,kagawa_dual-mode_2022}, but these cameras exhibit readout noise and dark current that drown out single-photon signals. On the other hand, charged-coupled device (CCD) and complementary metal oxide semiconductor (CMOS) cameras with single-photon sensitivity -- including electron multiplying CCD (EMCCD), intensified CCD (ICCD), scientific CMOS (sCMOS), and quantitative CMOS (qCMOS)~\cite{jerram_llccd_2001, ma_review_2022, seo_027e-rms_2015, roberts_comparison_2024} -- are widely used in quantum optics and astronomy. However, the frame rates of such cameras are typically well below 100~kfps, limited by the readout speed of the sensor. More exotic cameras, such as single-photon avalanche diode (SPAD) array cameras, provide time resolutions around 100~ps with exceptionally high readout rates~\cite{bruschini_single-photon_2019}, but low fill factors and pixel counts, combined with complexity in interfacing have limited widespread adoption. Focal plane photon counters (e.g., Lexitek PAPA~\cite{noauthor_lexitek_nodate, papaliolios_speckle_1985}) reach $10^6$~counts/sec with high spatial resolution, but are limited by the low quantum efficiency of the intensifier. Finally, superconducting nanowire single-photon detector (SNSPD) arrays~\cite{oripov_superconducting_2023, korzh_demonstration_2020, reddy_superconducting_2020} have both high quantum efficiency and fast time response, but require cryogenics and therefore a huge overhead in most applications that don't already operate in a cryogenic environment.

Given the robustness and widespread adoption of silicon technology, fast silicon-based cameras would be highly desirable. In this paper, we present a fast imaging scheme for EMCCDs and other cameras that rely on frame-transfer. At the hardware level, an EMCCD sensor is divided equally into an exposure and storage area, with the latter being physically masked to block any impinging light. To reach single-photon sensitivity, EMCCDs have an additional electron multiplying gain register before the output amplifier~\cite{jerram_llccd_2001} that reduces the input-referenced readout noise down to sub-electron level. In traditional EMCCD imaging, photoelectron charges accumulate in the exposure area; post exposure, these charges are shifted row by row into the storage area at vertical clock speed $f_\text{vert}$ of several megahertz. Afterwards, the charges are amplified and digitized row by row at a slower rate. The frame rate of EMCCDs is primarily limited by the slow readout time, as a new image exposure can start only after all charges from the last exposure have been read out. To overcome this issue, specifically designed camera chips with additionally inserted ``in-situ storage'' have been proposed~\cite{etoh_toward_2013}, for which the readout of all the data occurs only after the completion of video acquisition. However, none of the efficient signal amplification methods have been incorporated with those designs, which makes their noise level unsuitable for single photon detection and no widespread adoption.

\begin{figure*}[t] 
    \centering
    \includegraphics[width=\textwidth]{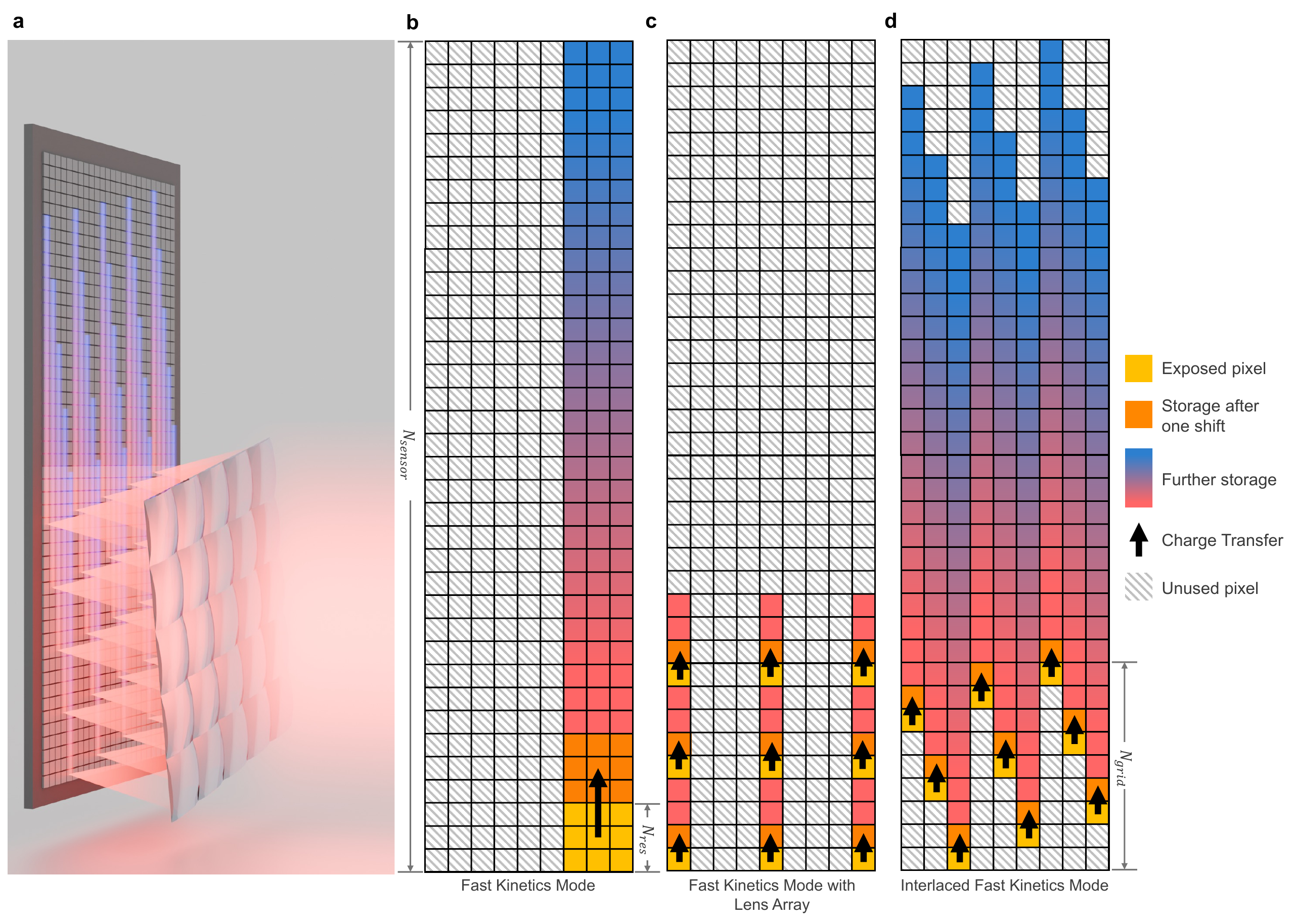}
    \caption{\textbf{Working principle of Interlaced Fast Kinetics Imaging}. \textbf{(a)} Schematic of our method: A lens-array is placed in front of the camera sensor to focus the input light (time dynamic depicted as a color gradient) onto the exposure pixels where individual columns capture the time dynamics during repeated row transfers. \textbf{(b)} In standard fast kinetics imaging, only a subsection of the sensor (yellow) is exposed. Before the next frame can be acquired the whole image has to be shifted out of the exposure area. Hence the effective frame rate is limited by the number of shifting operations necessary. \textbf{(c)} Introduction of a lens array enables us to spatially distribute the image across the sensor, so that only a single row transfer is necessary until the next frame can be captured. However, the number of frames that can be acquired is limited to few shifting operations and most sensor pixels are unused. \textbf{(d)} By precisely tilting the lens array, each exposed pixel is located in its own column, optimally filling the sensor. Interlaced fast kinetics mode allows both, a frame rate set by the row transfer rate and a maximum number of frames (See SI~\ref{SI:scalability}).}
	\label{fig:SetupFig}
\end{figure*}

Here, we introduce \emph{interlaced fast kinetics imaging} and experimentally demonstrate up to 3.33~Mfps burst video acquisition on a commercial single-photon sensitive EMCCD camera. The significant increase in frame rate is achieved by exposing the video \emph{during frame transfer}, and introducing a tilted lens array in the imaging path that focuses the input light onto a spatially separated grid of pixels, each located in its own column. During row transfer, the remaining unexposed pixels of each column act as intermediate ``in-situ storage'' for that pixel, ensuring that only one row shift per frame is necessary. After a maximum number of row shifting operations set by the sensor's resolution, the stored movie can be slowly read out with sub-electron sensitivity. In other words, photon arrival times are converted into spatial position on the EMCCD sensor; a high frame rate movie is then reconstructed from a single EMCCD ``image''. In the following, we introduce our concept and explain the necessary characteristics of the lens array, before presenting experimental benchmarks of the system. Finally, we present reconstructed 1.67~Mfps videos of a Gaussian beam rotating at a frequency of 488.3~kHz.

In order to appreciate the strengths of interlaced fast kinetics, we will begin by building an understanding of the state-of-the-art fast imaging mode on EMCCDs, called fast kinetics mode (FKM) (see SI~\ref{SI:Operatingmodes}). This mode, shown in Fig.~\ref{fig:SetupFig}b, performs a very simple trade-off of resolution for speed: By exposing only a small region of the sensor with height $N_{\text{res}}$, a new frame can be acquired as soon as the last image has been shifted out of the exposure area. This allows the temporary storage of many frames \emph{on the sensor} up to the point that the sensor is filled with stored frames. As such, the effective frame rate is determined by the time it takes to shift the image \emph{entirely} out of the exposure region of the sensor, given as $f_{\text{vert}}/N_{\text{res}}$. 

Our approach is faster by a factor of the image height: if we separate the exposed frame into a grid for which each grid point is exactly one pixel on the camera (see Fig.~\ref{fig:SetupFig}c), only one shifting operation is needed until the next frame can be captured. This drastically increases the frame rate to the row shifting rate $f_{\text{vert}}$. While the described grid could be generated using a mask or array of apertures (as proposed in~\cite{shiraga_ultrafast_1997}) with a substantial loss in signal photons, we instead utilize a lens array to maximize the photon signal. Unfortunately, in this configuration only few frames can be acquired before the shifted charges overlap with other exposed grid points, and most of the camera's pixels remain unused (see Fig.~\ref{fig:SetupFig}c).

This limitation can be overcome by slightly tilting the lens array and hence the grid (see Fig.~\ref{fig:SetupFig}d). When choosing appropriate spacings and angles between exposure pixels, such that each grid point occupies its own column, overlapping between adjacent grid points during shifting is fully avoided. This allows to drastically increase the number of frames, given by the number of sensor pixels in vertical direction minus the vertical size of the grid $N_{\text{sensor}}-N_\text{grid}$, which means interlaced fast kinetics mode optimizes the number of usable storage pixels. While similar ideas have been proposed~\cite{shiraga_ultrafast_1997, mandracchia_high-speed_2024,arai_back-side-illuminated_2013, etoh_toward_2013}, our method does not necessarily require custom designed camera chips (such as in-situ image storage CCDs) or custom optics, but instead utilizes commercial image sensors~\cite{daigle_characterization_2012} and lens arrays.

\section{Lens Array}
\label{sec:expsetup}
\begin{figure}[t] 
	\centering
    \includegraphics[width=\columnwidth]{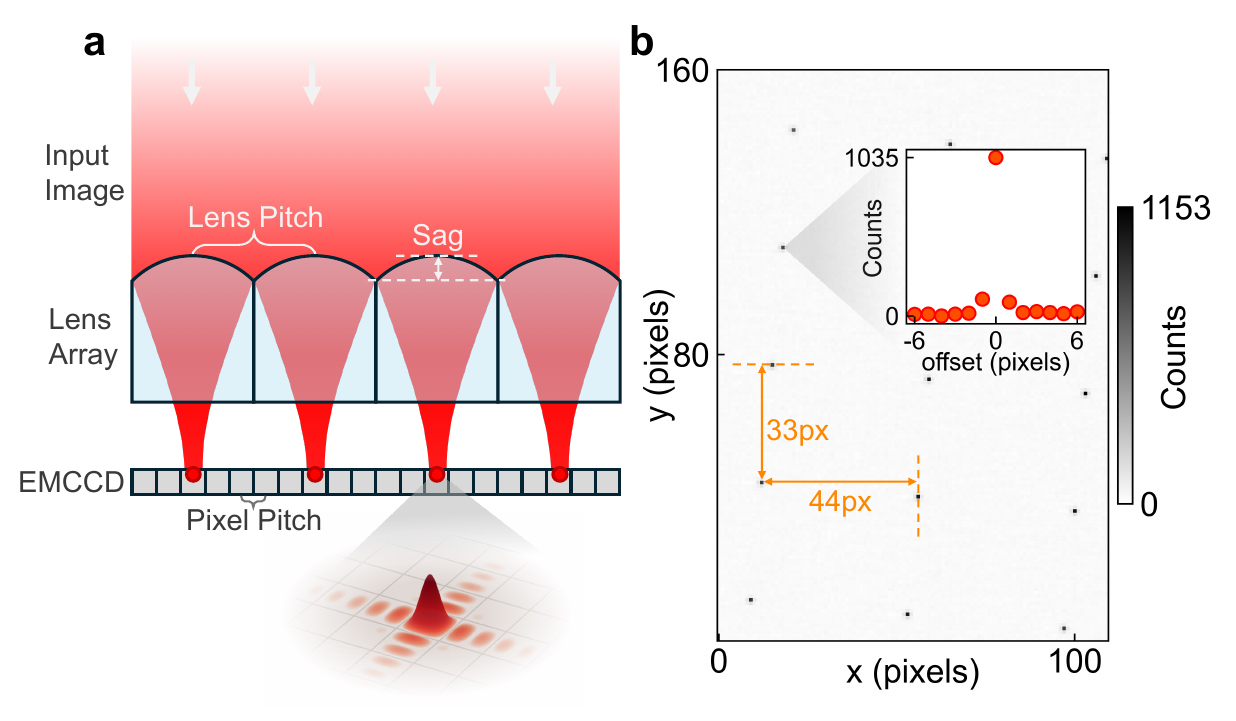}
	\caption{\textbf{Understanding properties of the lens array}. \textbf{(a)} Schematic of the focusing lens array. In order to reach maximum spatial and temporal resolution, the focus of a single lens in the array needs to be captured in a single camera pixel. Light diffracted beyond the edge of that pixel leads to spill-out \textbf{(inset)}. For a fully filled lens array, the separation between foci in units of focus size is proportional to $h/\lambda$ (Eq.~\ref{eq:N_p}) where $h$ is the sag of array and $\lambda$ is the wavelength of the light. \textbf{(b)} Experimentally observed pixel grid (after re-imaging optics), with a horizontal separation of 44~pixels and vertical separation of 33~pixels. \textbf{(inset)} shows a horizontal cross section, with $\sim$10\% leakage to neighboring pixels.}
	\label{fig:Sag}
\end{figure}

The fast imaging performance of our method crucially depends on the properties of the lens array. In particular, we require that the focal spot size of each individual lens must be smaller or comparable to the size of a single sensor pixel. Any leakage of light into vertical adjacent pixels will decrease temporal resolution, while horizontal leakage will introduce crosstalk between grid points. Furthermore, the separation between the foci must be large enough such that each grid point can be located in its own column. If focal waist and separation are incommensurate with the sensor dimensions, they can be rescaled by re-imaging (see SI~\ref{SI:Expsetup}). Hence, the key figure of merit is not the numerical aperture (NA) of each lens, but the separation of two grid points in units of the focal spot size, which we denote as $N_p$. Assuming each element in the lens array to be a plano-convex lens with approximately circular aperture, unity fill factor, refractive index $n$, and sag $h$, and furthermore assuming diffraction-limited conditions with a normally incident, collimated input beam of wavelength $\lambda$, we derive (see SI~\ref{SI:Distortion}):
\begin{equation}
    N_p=\frac{8(n-1)h}{2.44\lambda}
    \label{eq:N_p}
\end{equation}
Equation (\ref{eq:N_p}) shows that, for a given wavelength and lens materials, a larger sag, $h$, leads to higher $N_p$. In practice however, $N_p$ is limited by $h$. For example, $N_p=100$ requires a sag of $h$=47.5~$\mu$m for $\lambda=780$~nm and $n=1.5$, which is challenging to achieve with nanofabrication techniques. We therefore choose a lens array made by molding (Edmund Optics 72238), which provides $N_p$=347 and 195 in x \& y directions respectively. To ensure commensurability with the EMCCD sensor's (HNü 512) pixel grid, we use a commercial macro zoom lens to re-image the focal points right after the lens array onto the camera sensor (see SI~\ref{SI:Expsetup}), resulting in a 15$\times$11 grid of focal points. While we theoretically expect a spill-out within 1\% (corresponding to the average intensity of 4 adjacent pixels divided by the intensity of central pixel), experimentally we find the spill-out to be $\sim$10\% for the central grid points, limited by: (i) image side NA of the macro-lens; (ii) the higher order sinc-squared lobes of the lens aperture (Fig.~\ref{fig:Sag}a); and (iii) optical aberrations. We also characterize the nonlinear distortion of our re-imaging setup to be under 20\% relative to one pixel for most grid points (see SI~\ref{SI:Distortion}), and lower than one pixel for the edge grid points. We minimize the impact of this spill-out by leaving two free columns between each grid point. This completely eradicates spatial crosstalk at the cost of fewer pixel number in our movies (see SI~\ref{SI:scalability}). Spill-out remains in row-shift direction, manifesting as a slight reduction in temporal resolution. 

While we choose our lens array for a maximal separation $N_p$ to reduce spill-out, this reduces the overall photon collection efficiency to $\sim5$\% because its NA is significantly larger than the object side NA of the macro lens. This, however, can be straightforwardly increased to $\sim60$\% by selecting a lens array with smaller but enough $N_p$ and matching the NA with the macro lens. See SI~\ref{SI:Expsetup} for more discussion.

\section{Time Resolution Benchmarking}
\label{sec:contrastmeasure}
\begin{figure}[t] 
    \includegraphics[width=\columnwidth]{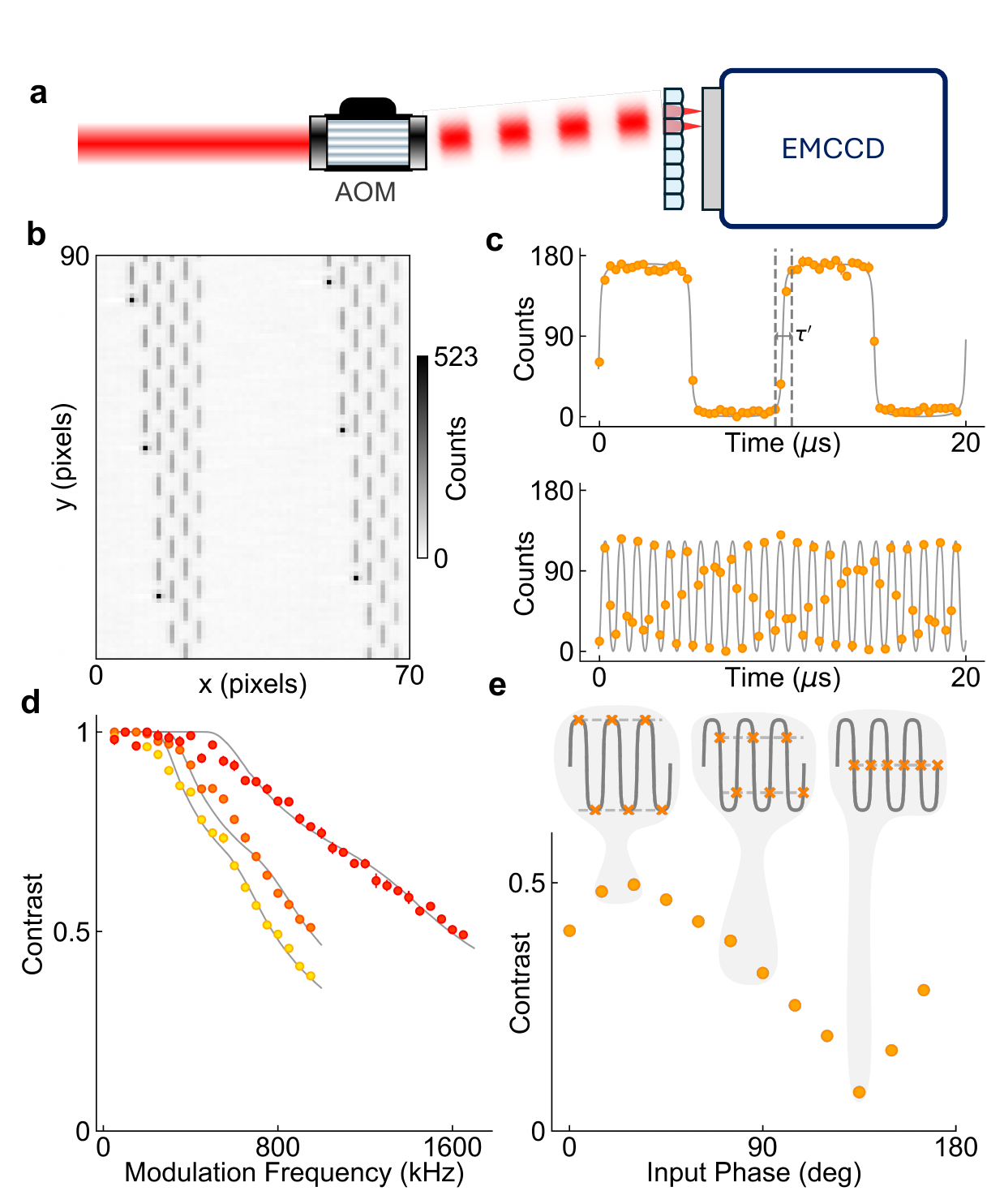}
    \caption{\textbf{Benchmarking temporal resolution}. \textbf{(a)} Schematic of the experiment. A time-varying signal with short rise and fall times is generated using an acousto-optical modulator. \textbf{(b)} Recorded raw image reveals the intensity modulation behavior (at 0.1 MHz) of the input beam as a dashing in the vertical (shifting) direction. \textbf{(c)} Extracted time-traces of a representative pixel for 0.1~MHz \textbf{(top)} and 1.15~MHz \textbf{(bottom)} at $f_{\text{vert}}=3.33$~MHz; the contrast drops as the modulation frequency approaches half of the row transfer frequency. The measured rise-time $\tau^{\prime}\approx0.9$~$\mu$s, is much slower than the AOM 10\% to 90\% rise-time of $50.2(6)$ ns. \textbf{(d)} Measured contrast vs optical modulation frequency for different $f_{\text{vert}}$ of 1.67~MHz (yellow, fast kinetics mode), 2~MHz (orange, normal mode) and 3.33~MHz (red, normal mode). Half-contrast occurs as modulation frequency approaches $f_{\text{vert}}/2$, indicating near-Nyquist performance. Each data point is extracted from a single pixel averaged over 10 shots and input signal is synced with camera trigger. The solid curves are models fits including effects of spill-out and finite row-shifting rate (see SI~\ref{SI:contrastbehavior}). \textbf{(e)} At commensurate frequencies (i.e., $f_{\text{vert}}$ at 2~MHz and input frequency at 1~MHz) a change in the relative phase between beam modulation and shifting leads to a change in contrast (as depicted schematically in the insets). All time-traces are background subtracted such that their minimum value is zero. }
    \label{fig:ContrastFig}
\end{figure}

Optimal time resolution for the interlaced fast kinetics camera is reached for perfect single-pixel focusing and instantaneous charge transfer. In this idealized limit the timing resolution is determined solely by the row-shift rate $f_{\text{vert}}$. In practice, the timing resolution is reduced by two factors: (i) vertical spill-out into adjacent camera pixels leading to temporal smearing at a fixed spatial location; (ii) non-instantaneous charge-transfer within the EMCCD itself, leading to temporal smearing (see SI~\ref{SI:contrastbehavior}). Given these factors we expect the frequency at which the contrast halves to be lower than the Nyquist frequency $f_{\text{vert}}/2$.
We benchmark the temporal response of our system by recording the camera's response when switching a large input beam on and off at several modulation frequencies (depicted schematically in Fig.~\ref{fig:ContrastFig}a). A representative region of the resulting raw image is shown in Fig.~\ref{fig:ContrastFig}b, where the on/off cycles are clearly visible in the vertical (shifting) direction as a dashed line. Note that we have left two empty columns between each grid point to suppress the impact of horizontal spill-out. We modulate the beam using an acousto-optic modulator (AOM; IntraAction ATM-1001A2) and optimize its temporal response by sending a tightly focused input beam ($\sim$200~$\mu m$ $1/e^2$ width) through the crystal, providing a measured 10\% to 90\% rise time of 50.2(6)~ns (measured on a photodiode), allowing us to create step-like input signals to our camera. Significant deviations from a step-like behavior due to imperfect input is therefore only expected at intensity modulation frequencies around 10~MHz, about an order of magnitude faster than the camera's dynamics. Fig.~\ref{fig:ContrastFig}c depicts traces for one grid point extracted from an image analogous to Fig.~\ref{fig:ContrastFig}b, but at 0.1~MHz (top) and 1.15~MHz (bottom) intensity modulation frequency. Even for the low frequency modulation, the step can be wider than one time bin due to temporal broadening caused by spill-out and charge transfer dynamics. At highest modulation frequencies, we observe a contrast reduction, as the on-time of the beam decreases toward the exposure time per frame (inverse of row transfer rate).

To quantify the frequency response, we compute the contrast for a fixed grid point as a function of the intensity modulation frequency. The results are shown in Fig.~\ref{fig:ContrastFig}d. We repeat this measurement using several operating modes of the camera, differing primarily by their row shifting rate of 1.67~MHz (yellow line, fast kinetics mode), 2~MHz (orange line, normal mode) and 3.33~MHz (red line, normal mode). Details about the camera operating modes can be found in SI~\ref{SI:Operatingmodes}. For all three cases we observe excellent contrast at low frequencies that decreases by half as the modulation frequency approaches the Nyquist limit (given by half the row-transfer frequency) at i. 975(30)~kHz, ii. 785(8)~kHz, iii. 1619(25)~kHz. Solid lines in Fig.~\ref{fig:ContrastFig}d represent model fits, using the charge shifting time as the fitting parameters. Further information on the theoretical model can be found in SI~\ref{SI:contrastbehavior}. 

At an input frequency of half the shifting frequency, we can directly observe the impact on the contrast of the phase delay between the row-shift and the intensity modulation. This behavior is shown in Fig.~\ref{fig:ContrastFig}e: maximum contrast is achieved for matching phases (insets in Fig.~\ref{fig:ContrastFig}e) since the charge transfer then coincides temporally with the beam turn-on. The contrast decreases as the phase offset grows, dropping to zero when shifting occurs at the precise center of the on/off intervals. For incommensurate frequencies (as used in Fig.~\ref{fig:ContrastFig}d), the phase offset varies over modulation cycles so we are able to extract the maximum contrast for Fig.~\ref{fig:ContrastFig}d.

\section{Video Reconstruction}
\label{sec:video}
\begin{figure}[t] 
    \includegraphics[width=\columnwidth]{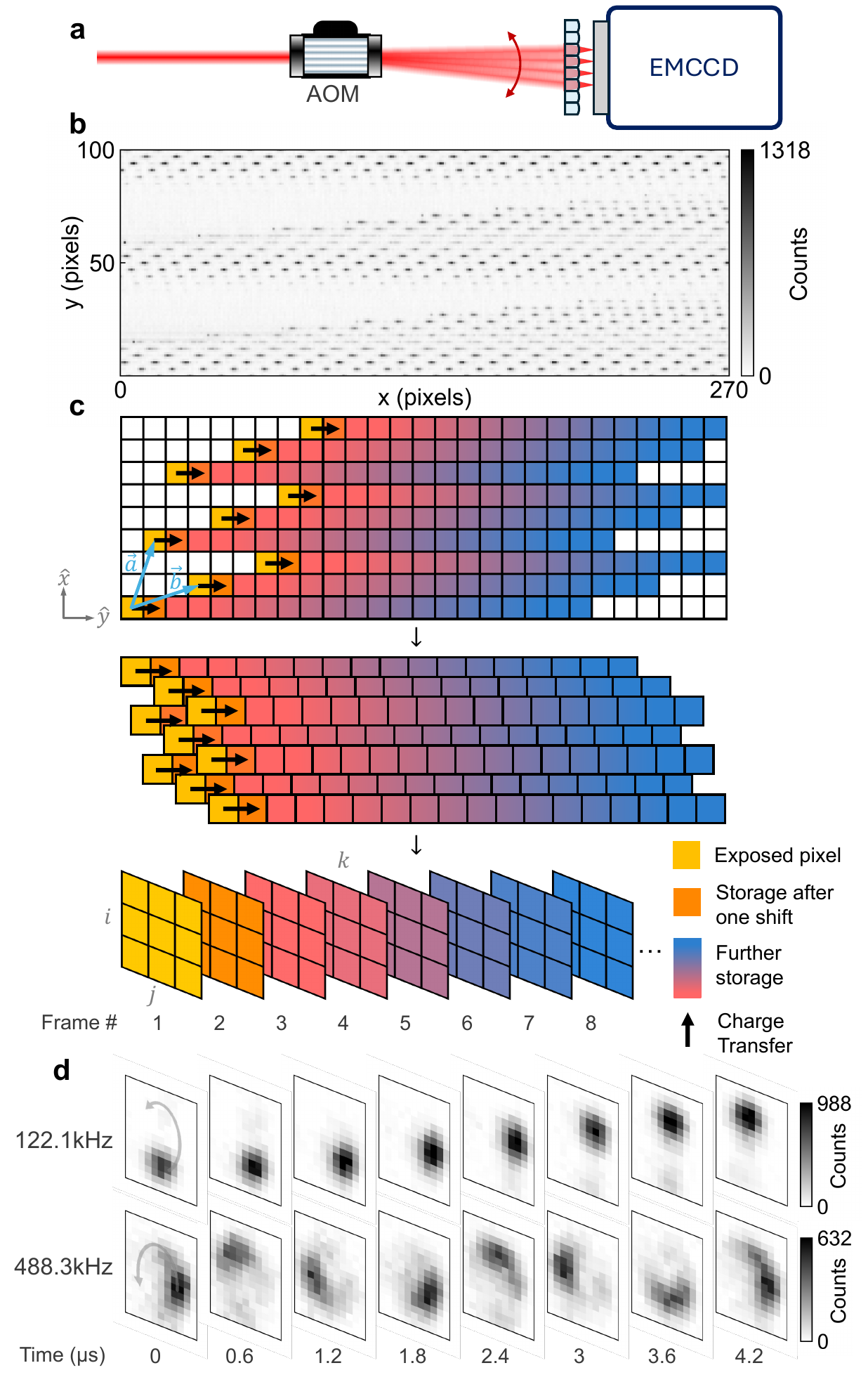}
    \caption{
        \textbf{Video reconstruction}. \textbf{(a)} Schematic of the setup. Two AOMs (only one shown) steer a laser spot around a circle which is focused onto the fast camera. \textbf{(b)} Captured raw image from which the video is reconstructed. Image is rotated by 90 degrees compared to earlier figures. \textbf{(c)} Reconstruction process: Top row: Each image row corresponds to a unique spatial location in the image, with the column reflecting the time. Middle row: arranging the rows into planes of constant time produces the array of frames shown in the (\textbf{bottom row}). $\vec{a}$, $\vec{b}$ represents two primitive vectors of the grid. $\hat{x}$, $\hat{y}$ denotes two orthogonal directions along the column and row directions of the sensor. $i,j$ denotes two spatial indices and $k$ represents the time index. \textbf{(d)} Eight consecutive frames from the reconstructed videos taken at 1.67~Mfps. The spot's rotation frequency is 122.1~kHz (top) and 488.3~kHz (bottom).}
	\label{fig:FMFig}
\end{figure}

Finally, we showcase the ability of interlaced fast kinetics imaging to take high frame rate, two-dimensional movies. To create a fast moving scene, we employ two crossed AOMs (IntraAction ATM-1001A2) to steer a spot in 2D, generating a circular motion with a rotation frequency tunable within 1~MHz (see also Fig.~\ref{fig:FMFig}a, SI~\ref{SI:Expsetup}). The number of frames that we can collect is limited by the difference between the sensor height and the vertical size of the grid (504 pixels). In order to exploit the whole sensor size including the storage area, we nominally run the camera in FKM mode using a ROI (region of interest) with single pixel height, but exposure according to Fig.~\ref{fig:SetupFig}d. In total this allows up to 544 frames (see SI~\ref{SI:Operatingmodes}). A larger sensor would increase the number of frames further (see SI~\ref{SI:scalability}).

We take videos at frame rate of 1.67~Mfps with 540 frames, resulting in a total video length of 324~$\mu$s. A typical read-out image, from which a video can be extracted, is shown in Fig.~\ref{fig:FMFig}b. The extraction procedure is shown in Fig.~\ref{fig:FMFig}c. In short, we map each pixel of the 2D image of the camera $I_c(\vec{r})$ onto a space-time position in the movie $I_m(i,j,\tau_k)$, where $\vec{r}$ represents the position on camera pixels, $i,j$ represent spatial indices for movie frames, and $\tau_k=k/f_{\text{vert}}$ is the discrete time for frame $k$. Defining $r_0$ as the origin point of the grid, $\vec{a}$, $\vec{b}$ as two primitive vectors (see Fig.~\ref{fig:FMFig}c), we reconstruct $I_m$ according to
\begin{equation}
I_m\left(i,j,\tau_k\right)=I_c\left(r_0+i\vec{a}+j\vec{b}+k\hat{y}\right)
\end{equation} 
Finally, we arrange the frames into a video. In Fig.~\ref{fig:FMFig}d we show eight consecutive frames of a movie that captures the circular motion of a laser beam orbiting with frequencies of 122.1 and 488.3 kHz. The full videos can be found as Supplementary Movies 1 and 2. 

\section{Summary \& Outlook}
\label{sec:outlook}

In summary, we have developed a fast imaging technique for EMCCD cameras by leveraging a precisely tilted lens array for in-situ storage of individual pixels, enabling burst video acquisition at up to 3.33~Mfps. We demonstrated the ability to maintain 50\% contrast for intensity modulation at frequencies up to 1.61 MHz, near the Nyquist limit set by the camera's row shifting rate. Furthermore, we captured 1.67~Mfps video with 540 frames at a resolution of 11×15 pixels. The setup demonstrated in this work provides $\sim5$\% light collection efficiency at 780 nm; straightforward changes to the optical setup can increase its efficiency to 60\%, doubling the sensor size would allow $\sim100$ frames at image resolutions of $50\times50$ (see SI~\ref{SI:Expsetup} and~\ref{SI:scalability}) .

Interlaced fast kinetics imaging has \emph{immediate applications} for a variety of imaging problems benefiting from high frame rate while requiring single-photon sensitivity: in biology the camera can either serve as an many-pixel photon counter in correlation based imaging~\cite{buchholz_widefield_2018, israel_quantum_2017, schwartz_improved_2012} or directly facilitate fast fluorescence imaging in high throughput flow cytometry, calcium dynamics tracking and video-rate nanoscopy~\cite{rees_imaging_2022, kumar_high-speed_2011, krishnaswami_towards_2014}; detection of individual photons in photonic quantum matter would enable extraction of high-order correlation function~\cite{clark_observation_2020}; in quantum computing \& simulation, fast imaging will enable mid-circuit measurements for error correction~\cite{chamberland_fault-tolerant_2018} and unraveling of random quantum circuits~\cite{ skoric_parallel_2023}. 

\section{Acknowledgments}
This work was supported by AFOSR MURI FA9550-19-1-0399 and NSF QLCI-HQAN 2016136.

\section{Author Contributions}
The experiments were designed by M.G., J.S., B.L., L.P., M.J. and performed by B.L.. The apparatus was built by B.L. and Y.C.F.. The project was supervised by J.S.. All authors analyzed the data and contributed to the manuscript.

\section{Competing Interests}
The authors declare no competing financial or nonfinancial interests.

\clearpage


\putbib
\end{bibunit}


\subsection{Data Availability}
The experimental data presented in this manuscript are available from the corresponding author upon request, due to the proprietary file formats employed in the data collection process.
\subsection{Code Availability}
The source code for simulations throughout are available from the corresponding author upon request. 
\subsection{Additional Information}
Correspondence and requests for materials should be addressed to B.L. (bl254@stanford.edu). Supplementary information is available for this paper.

\clearpage
\newpage




\begin{bibunit}
\onecolumngrid
\newpage
\section*{Supplementary Information}
\appendix
\renewcommand{\appendixname}{Supplement}
\renewcommand{\theequation}{S\arabic{equation}}
\renewcommand{\thefigure}{S\arabic{figure}}
\renewcommand{\figurename}{Supplemental Information Fig.}
\renewcommand{\tablename}{Table}
\setcounter{figure}{0}
\setcounter{table}{0}
\numberwithin{equation}{section}

\section{Experimental Setup}
\label{SI:Expsetup}

\begin{figure*}[ht] 
	\centering
 	\includegraphics[width=0.75\textwidth]{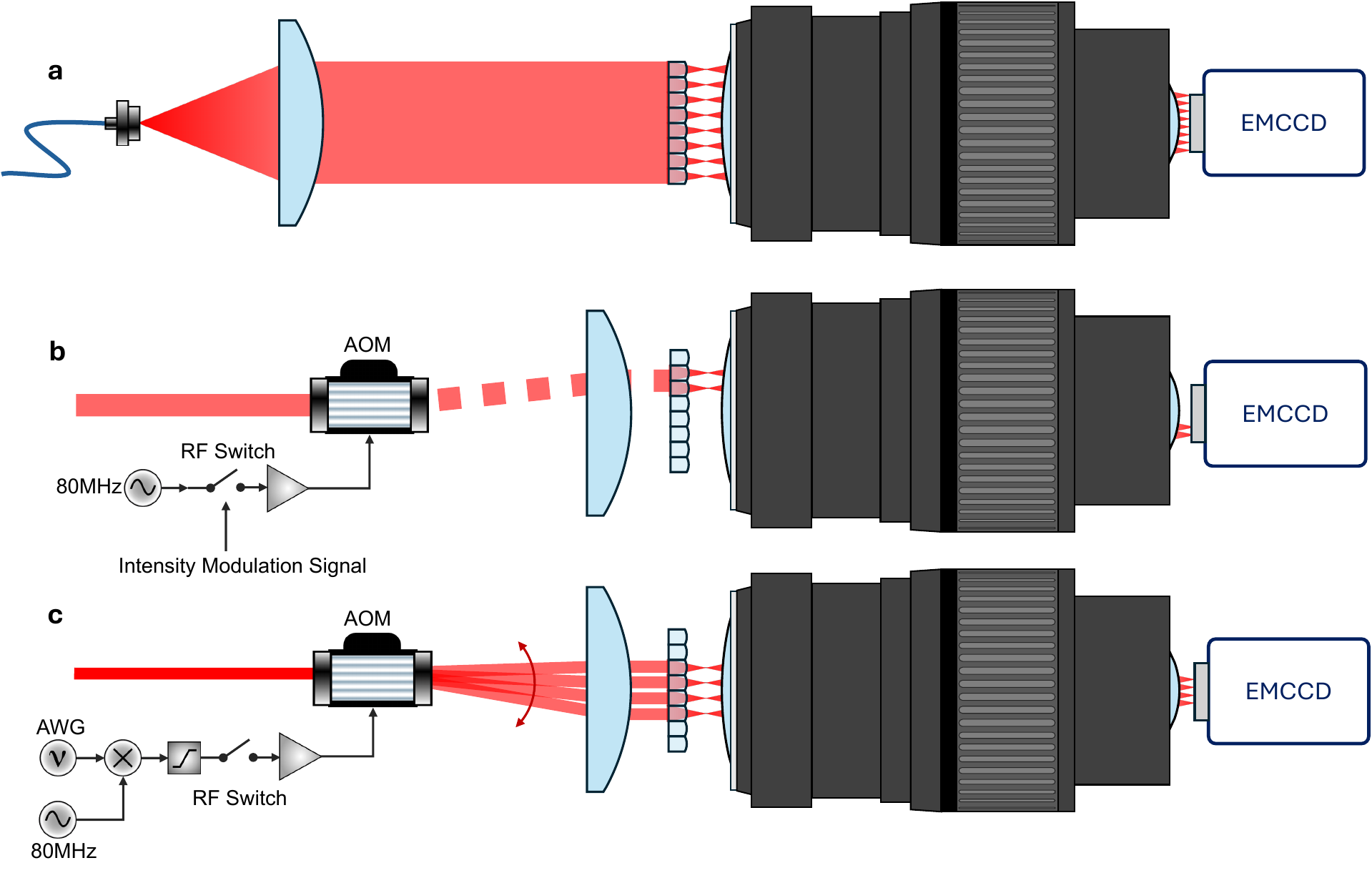} 
	\caption{
		\textbf{Schematic of experimental setup}. Since the focal length of the lenslet array  (Edmund Optics 72238) is shorter than the flange distance of the EMCCD camera, we use a commercial macro lens (Canon MP-E 65~mm f/2.8 1–5x Macro) in reverse configuration to re-image the lens array onto the camera \textbf{(a)} Calibration setup. A collimated laser beam fully illuminates the lens array. This configuration is used to analyze and minimize alignment errors (see SI~\ref{SI:Distortion}) \textbf{(b)} Time contrast measurement setup. A square wave intensity modulation (50\% duty cycle) is generated using an RF switch and an acousto-optic modulator (AOM). \textbf{(c)} Video acquisition setup. Sine wave signals (generated by Redpitaya) with sinusoidal frequency sweeps between 10~MHz and 50~MHz are up-converted to a frequency sweep in the range of 90~MHz to 130~MHz by mixing with an 80~MHz local oscillator. The sweep rate is tunable within 1~MHz and the relative sweeping phase is 90 $\deg$ between two channels. These signals drive two crossed AOMs to create rotating pattern in 2D (only one AOM shown).
	}
	\label{fig:expsetup}
\end{figure*}

In order to achieve high frame-rate imaging we need to make the grid of focal points stemming from the lens array commensurate with the pixel grid of the camera sensor. Using a lens array to directly image onto the sensor is not possible for two reasons: (1) We were unable to find a lens array with exactly the right distances between the lenses; and (2), the sensor of our EMCCD camera is located behind a 17.75-mm-long flange which is longer than the focal length of our chosen lenslet array. Hence, an additional re-imaging system is needed. A 4f re-imaging system with a large field of view that allows maintaining a diffraction limited focus across the whole array is challenging and cannot be achieved with only two spherical lenses. Instead, we utilize a commercial macro zoom lens with adjustable magnification (Canon MP-E 65~mm f/2.8 1–5x Macro) to re-image the focal points onto the sensor (Fig.~\ref{fig:expsetup}). Due to the large size of our lenslet array (Edmund Optics 72238; 17.6$\times$18~mm),  we use the macro lens in reverse configuration (de-magnifying) in which it supports a field of view as large as 36$\times$24~mm. Since the magnification ratio of macro lens is tunable from $1\times$ to $5\times$ when used normally, by reversely using the lens, we have a de-magnification ratio of 0.2-1$\times$, which makes it possible to image the whole lenslet array onto the EMCCD sensor with size of 8.19$\times$8.19~mm. 

\begin{figure}[ht]
	\centering
 	\includegraphics[width=\textwidth]{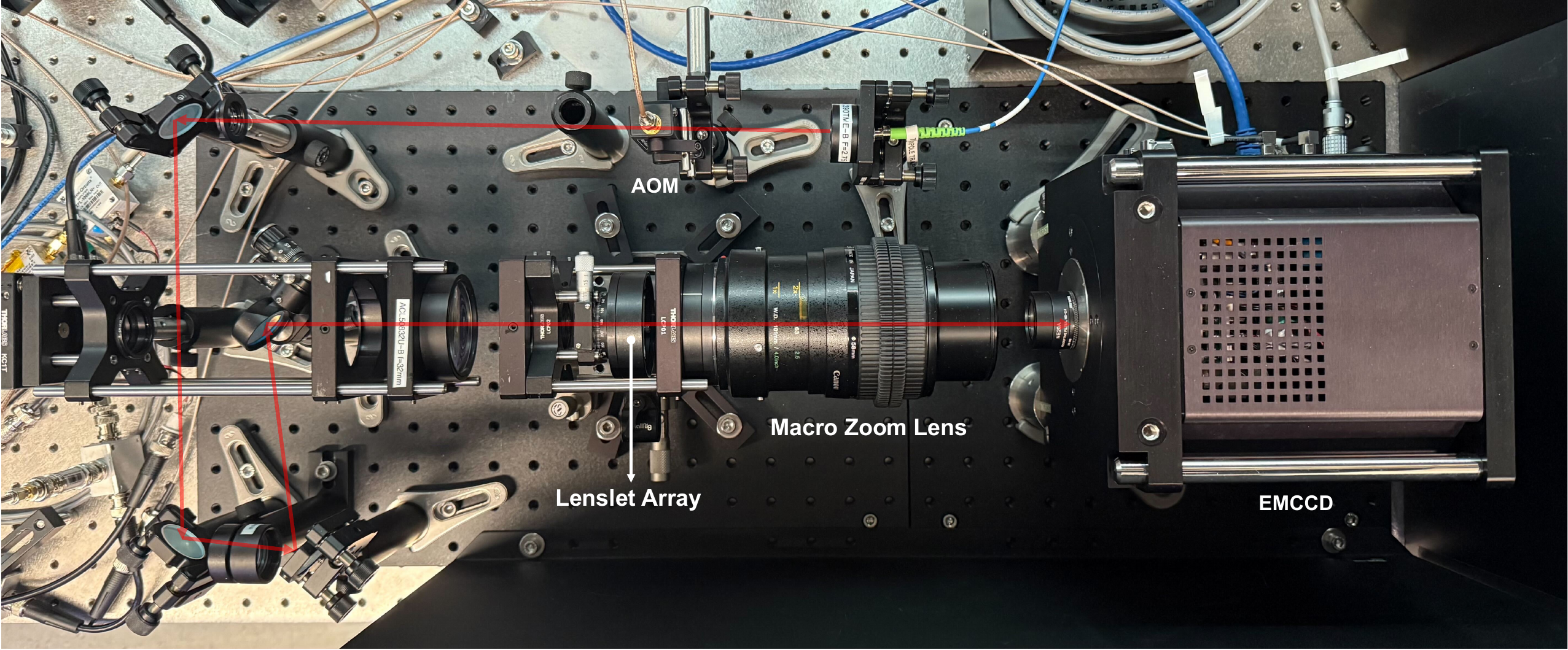}
	\caption{\textbf{Photo of experimental setup.} The commercial macro zoom lens is shown to be used reversely to image the lens array onto the EMCCD sensor. The lenslet array is placed in front of the macro zoom lens. The macro zoom lens is mounted together with the lenslet array on a 3-axis translation stage to provide enough degrees of freedom for fine alignment (see SI~\ref{SI:Distortion}). We set up the AOM aside the lenslet array which either modulate the intensity of input signal or deflect the beam in 2 directions (Only one AOM is installed in this photo). Beam path is depicted as red arrows in the photo.}
	\label{fig:Photo of experimental setup}
\end{figure} 

Schematics of our experimental configurations are shown in Fig.~\ref{fig:expsetup}. For calibration (see SI~\ref{SI:Distortion}) the setup is arranged as in Fig.~\ref{fig:expsetup}(a) with a large collimated beam that illuminates the full lenslet array. The focal points are then re-imaged on the EMCCD sensor using the macro lens. Note that in the setups we assume telecentric input before the lens array.

For contrast measurement (as shown in Fig.~\ref{fig:ContrastFig}) we send a constant tone of 80~MHz to an acousto-optic modulator (AOM) and turn this signal on and off using an RF switch (ZYSWA-2-50DR+). We measure a 10\% to 90\% rise time of the AOM on a photodiode to be 50.2(6)~ns for our input beam  ($\sim$200~$\mu m$ $1/e^2$ width).  A photo of our exerpimental setup is shown in Fig.~\ref{fig:Photo of experimental setup}.

For video acquisition benchmarking in 2D we first generate two frequency modulated signals using two output channels from a Redpitaya. We sinusoidally modulate the frequency of signals such that the frequency in each channel is swept in the range from 10 to 50~MHz with the sweeping rates tunable within 1~MHz. We then mix it with an 80~MHz local oscillator (using a Fluke 6060B) and up-convert it into the range of 90-130~MHz to fit with the working frequency of our AOMs. For creating circular rotating pattern we set the relative sweeping phase as 90 $\deg$ between the two channels while keeping the sweeping rates to be the same. We then send the signals to two crossed AOMs (IntraAction ATM-1001A2) to deflect the beam in 2D on the image plane. Finally the beam passes through the lens array and is re-imaged onto the camera. The video reconstruction is processed as described in the main text.

Our platform admits a variety of improvements in terms of image quality: re-imaging optics with a higher image-side NA would reduce spill-out and increase spatial and temporal resolution; extending the technique to multiple cameras enables continuous imaging. It is worth mentioning that the setup demonstrated in this work provides $\sim5$\% light collection efficiency at 780 nm; this can be increased to $\sim60$\% by either increasing the object side NA of re-imaging lens or reducing the output NA of lens array while maintaining the number of separation at a usable level. The collection efficiency can even be further increased by properly coating the re-imaging lens or integrating a lens array directly in front of the camera sensor to obviate the need for re-imaging optics. 

\section{Description of EMCCD operating modes}
\label{SI:Operatingmodes}
\begin{figure}[ht]
	\centering
 	\includegraphics[width=\textwidth]{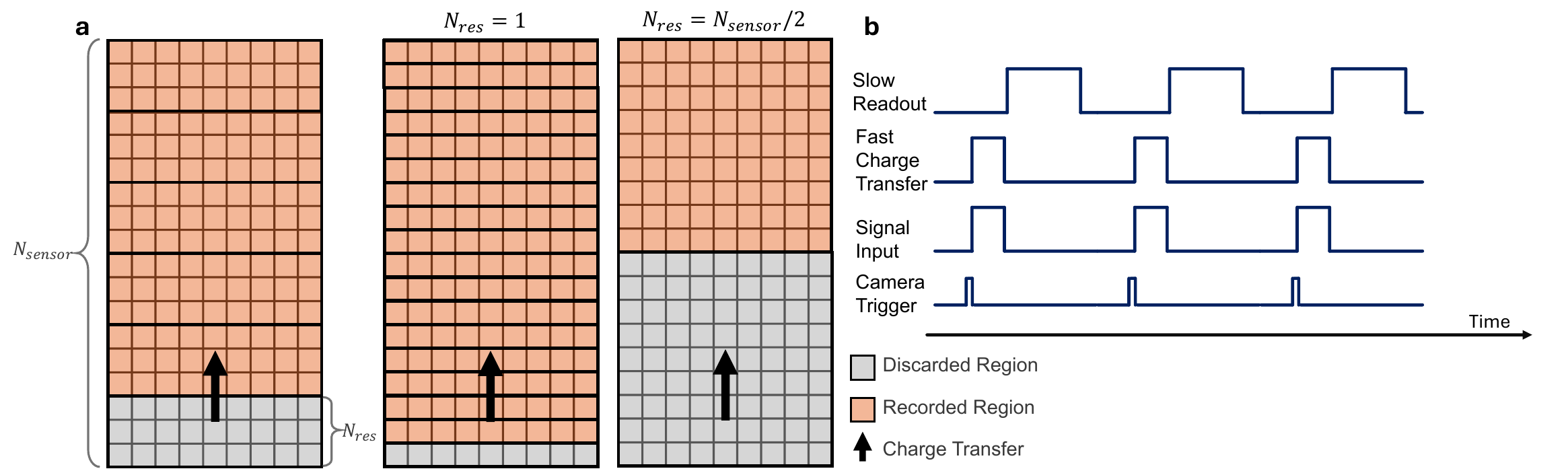}
	\caption{\textbf{Description of EMCCD fast kinetics mode (FKM)}. \textbf{(a)} Schematics of fast kinetics mode. \textbf{(left)} In fast kinetics mode a subsequent exposure is taken after the charges are shifted out of the exposed region, increasing the frame rate to $f_{\text{vert}}/N_{res}$. And the maximum number of frames is given by $\left\lfloor N_{\text{sensor}}/N_{\text{res}} \right \rfloor-1$. \textbf{(middle)} By setting $N_{\text{res}}=1$ we acquire both the maximum number of frames, $N_{\text{sensor}}-1$, and make the most use of the sensor area. \textbf{(right)} The fast kinetics mode will be nominally equivalent to normal mode when the ROI is set to $N_{\text{res}}=N_{\text{sensor}}/2$. \textbf{(b)} Time sequence of the experiment. We externally trigger the camera and sync charge transfer with the input light signal to ensure the recorded region isn’t over-exposed during slow readout.}
	\label{fig:EMCCD operating modes}
\end{figure} 

The specific EMCCD model, that we use in this work, is a N\"uv\"u HN\"u 512 based on a teledyne CCD 97 sensor. The sensor is a frame transfer CCD that possesses both exposure and storage area, with a physical mask on top of the storage area to avoid its exposure. Each area takes about half of the sensor size (512x512), and the total usable sensor size is $1048\times512$\cite{noauthor_ccd97_nodate}. Note that, there are 24 additional pixels between the exposure and the storage area. 

Our experiment makes use of two EMCCD operating modes: "normal" mode and "fast kinetics" mode (FKM). In normal mode the exposure area is first exposed once, before all the charges are transferred into the storage area where they are shielded from further exposure. After the (slow) readout, a new exposure can start. Hence, in this mode, the maximum image size that can be read out is 512x512. In fast kinetics mode, a subsequent exposure is taken right after the charges from the previous exposure are entirely shifted out of region of interest (ROI; with size $N_{res}$) and before the readout occurs. Once the sensor is filled, the slow readout occurs. This increases the burst video frame rate to $f_{\text{vert}}/N_{res}$ with a maximum frame number of $\left\lfloor N_{\text{sensor}}/N_{\text{res}} \right \rfloor-1$, where the last frame is not counted as it might be over-exposed during the readout. Note that the fast kinetics mode is nominally equivalent to the normal mode for a ROI covering the full exposure area, $N_{\text{res}}=N_{\text{sensor}}/2$ (Figure~\ref{fig:EMCCD operating modes}a).

For a readout of the whole sensor, we use FKM with an ROI of $N_{\text{res}}=1$. This allows to acquire both the maximum number of frames as $N_{\text{sensor}}-1$ and make the most use of the sensor area. In practice we define the ROI as 512 columns by 1 row at the bottom of the sensor (readout occurs at the top of the sensor) leading to a read-out image size of $1047\times512$.

Furthermore, in order to avoid overexposure during the slow readout phase (51.2~$\mu$s or 25.6~$\mu$s for reading out an entire row, for a horizontal shifting frequency of 10~MHz or 20~MHz, respectively), we switch off the input beam after the exposure. This is achieved by externally triggering the acquisition of the camera with a signal that is synced with the input light signal as shown in \ref{fig:EMCCD operating modes}b.

\section{Single pixel alignment and distortion measurement}
\label{SI:Distortion}

\begin{figure*}[ht!] 
	\centering
 	\includegraphics[width=0.5\textwidth]{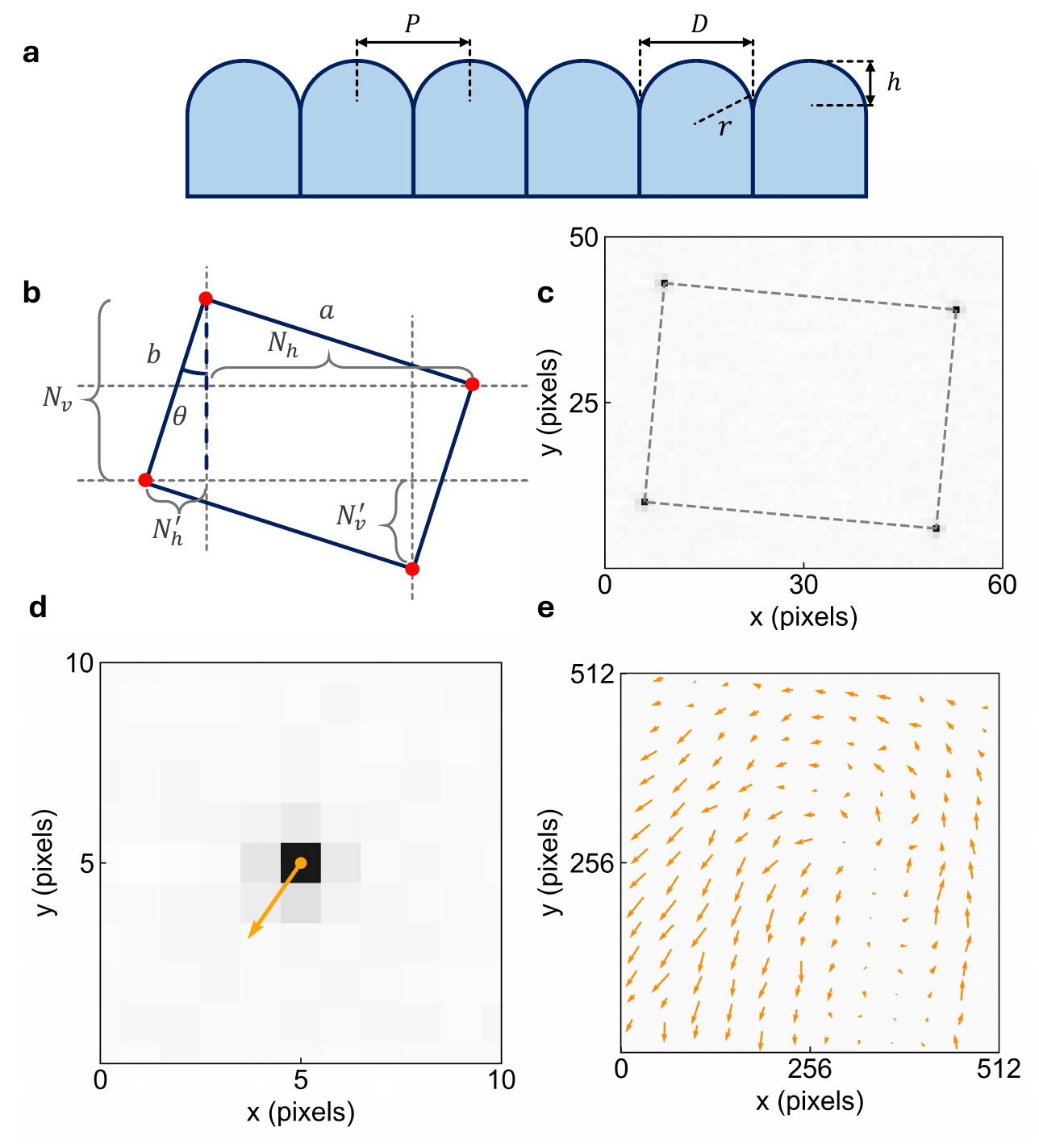}
	\caption{\textbf{Demonstration of fine alignment onto single pixel}. \textbf{(a)} Sketch of the relevant lens array geometry. We denote $P$ as the pitch of the lens array, $D$ as the size of the lens elements, $r$ as the radius of curvature of the spherical lens surface, and $h$ as the sag. \textbf{(b,c)} For lenslet arrays that are not necessarily square, certain relationship between $\theta$ and $a/b$ needs to be satisfied. \textbf{(c)} shows a real image of our grid points. For our case with $a/b=3/4$, we have $\theta$=5.19° and $N_h=4k, N_h^{\prime}=3k', k,k'\in \mathbb{N}$, which poses a relatively strict requirement for the alignment.  (d) Alignment error of a single pixel. The vector shows the centroid of intensity distribution relative to the pixel. \textbf{(e)} Linear alignment errors of the collective grid (translation, rotation, zoom) are calculated from the vector-field of individual grid points. The vector length scales in \textbf{(d,e) }are exaggerated in the plots.}
	\label{fig:distortion1}
\end{figure*} 
As outlined in the main text, we need to focus the input light onto spatially separated individual pixels. The lens array's ability to create such a grid with enough spacing is determined by the sag of the lens array as we will show in the following. We assume each element in the lens array to be a plano-convex lens with approximately circular aperture, diameter $D$, and focal length $f$. Under diffraction-limited conditions, when the collimated beam with wavelength $\lambda$ is incident normally on the lens, the focal waist in the imaging plane is given by the diameter of Airy disk's first zero $D_{\text{Airy}}=2.44 \lambda f /D$. Denoting the spacing between two individual lenses as pitch $P$ (Fig.~\ref{fig:distortion1}a), we find
\begin{equation}
    N_p\approx\frac{P}{D_{\text{Airy}}}=\frac{PD}{2.44 \lambda f}=\frac{P}{2.44 \lambda}\cdot 2\text{NA},
    \label{eq:diffratio}
\end{equation}
where we have introduced the numerical aperture in air as $\text{NA}\approx D/2f$. Hence, for a fixed pitch, $N_p$ will scale linearly with the $\text{NA}$ of the lens elements and generally a larger $\text{NA}$ is preferred. For maximum light collection, we aim for unit filling, $D=P$, which gives us:
\begin{equation}
    N_p=\frac{D^2}{2.44\lambda f}
\end{equation}
Using the lensmaker's equation $f=r/(n-1)$, where $r$ denotes the radius of curvature of the individual lens and $n$ the refraction index of the lens material, and furthermore noticing that the sag $h$ (Fig.~\ref{fig:distortion1}a) of a spherical surface is defined as $h=r-\sqrt{r^2-(D/2)^2}\approx \frac{D^2}{8r}$, with approximation assumes $D\ll r$, we find
\begin{equation}
    N_p=\frac{8(n-1)h}{2.44\lambda}
\end{equation}
indicating that the separation between grid points is given by the sag in optical wavelengths. In practice, this quantity is typically limited by the lens fabrication method's ability to create deep optics, i.e., large sag.

Our rectangular lenslet array (Edmund Optics 72238), provides 15$\times$11 lens array with pitch $P$ of 1.6~mm$\times$1.2~mm, effective focal length of $f=4.64$~mm, and sag of $h=144$~$\mu$m, for which we calculate $N_p$=347 and 195, with a focal spot sizes of 4.6 and 6.1~$\mu$m in x \& y directions respectively. As the pixel size of the camera is 16~$\mu$m x 16~$\mu$m, we theoretically expect a spill-out within 1\% (corresponding to the average intensity of 4 adjacent pixels divided by the intensity of central pixel). Experimentally we find the spill-out to be $\sim$10\% for the central grid point, limited by: (i) image side NA of the macro-lens; (ii) the higher order sinc-squared lobes of the lens aperture; and (iii) optical aberrations from both the lenslet array and the macro lens.

After creating the grid of foci, the precise alignment of the grid points onto commensurate single pixels on EMCCD is critical to both spatial and time resolution of burst video acquisition. Denoting distances inside each individual cell of the grid as shown in Fig.~\ref{fig:distortion1}b, where $a$, $b$ are the pitches of the rectangular lens array, and denoting $R$ as the magnification ratio of the re-imaging system, $\theta$ as the rotation angle and $p$ as the size of a single camera pixel, we arrive at the following set of equations:
\begin{equation}
Ra\cos\theta=N_h p, \quad 
Rb\sin\theta=N^{\prime}_h p,\quad
Rb\cos\theta =N_v p,\quad
Ra\sin\theta=N_v^{\prime}p
\end{equation}
where $N_h,N_h',N_v,N_v'\in \mathbb{N}$. By solving this set of equations we find:
\begin{equation}
    \label{eq:singlealignment}
    R=\frac{p}{ab} \sqrt[]{a^2{N^{\prime}_h}^2+b^2N_h^2},\quad \tan\theta=\frac{aN_h^{\prime}}{bN_h}, \quad N_v=\frac{b}{a}N_h,\quad N_v^{\prime}=\frac{a}{b}N_h^{\prime}
\end{equation}
In our case, $a/b=4/3$, we find that according to~\ref{eq:singlealignment}, $N_h$ can only be integer factors of 4 and $N_h^{\prime}$ can only be integer factors of 3. In our experiment we choose $N_h=44$ and $N_h^{\prime}$=3, which corresponds to $1/R=2.263$ and $\theta=5.194\deg$, posing a strict requirement for the alignment.

To make sure our setup is capable of aligning to the above set of parameters as close as possible, we first mount the lenslet array on a high-precision rotation mount (Thorlabs CRM1PT), which has enough precision to reach the desired $\theta$ within 5~arcmin. In order to get enough translational degrees of freedom, we mount both the macro zoom lens and the lenslet array together on a Newport 562-XYZ 3-axis translation stage, which provides the translation degrees of freedom in horizontal and vertical direction. The third translation degree of freedom of the stage is used to focus the output beam from the macro lens onto the sensor. 

For alignment and distortion measurements we collimate the fiber output beam to fully illuminate the lens array (Fig.~\ref{fig:expsetup}). We start by turning the rotation mount to the computed $\theta$. Then we adjust the zoom ratio of the macro lens to the desired $R$, while simultaneously adjusting the focus, to reach a course single pixel alignment for the grid points. 

After this coarse alignment, significant residual errors remain. Our system allows to correct for linear alignment errors, which are translations, zoom ratio and rotation angle of lens array, but any nonlinear distortion of the position on the grid point (which we cannot correct for using aforementioned linear alignment operations) must be smaller than one pixel. In particular this applies to the marginal edges. As it is difficult to reach a precise alignment by only observing the pixel grid on a camera, we compute the linear alignment errors from the constructed misalignment vector fields. In the following we describe our way of quantitatively benchmarking the alignment error.

\begin{figure*}[t] 
	\centering
 	\includegraphics[width=\textwidth]{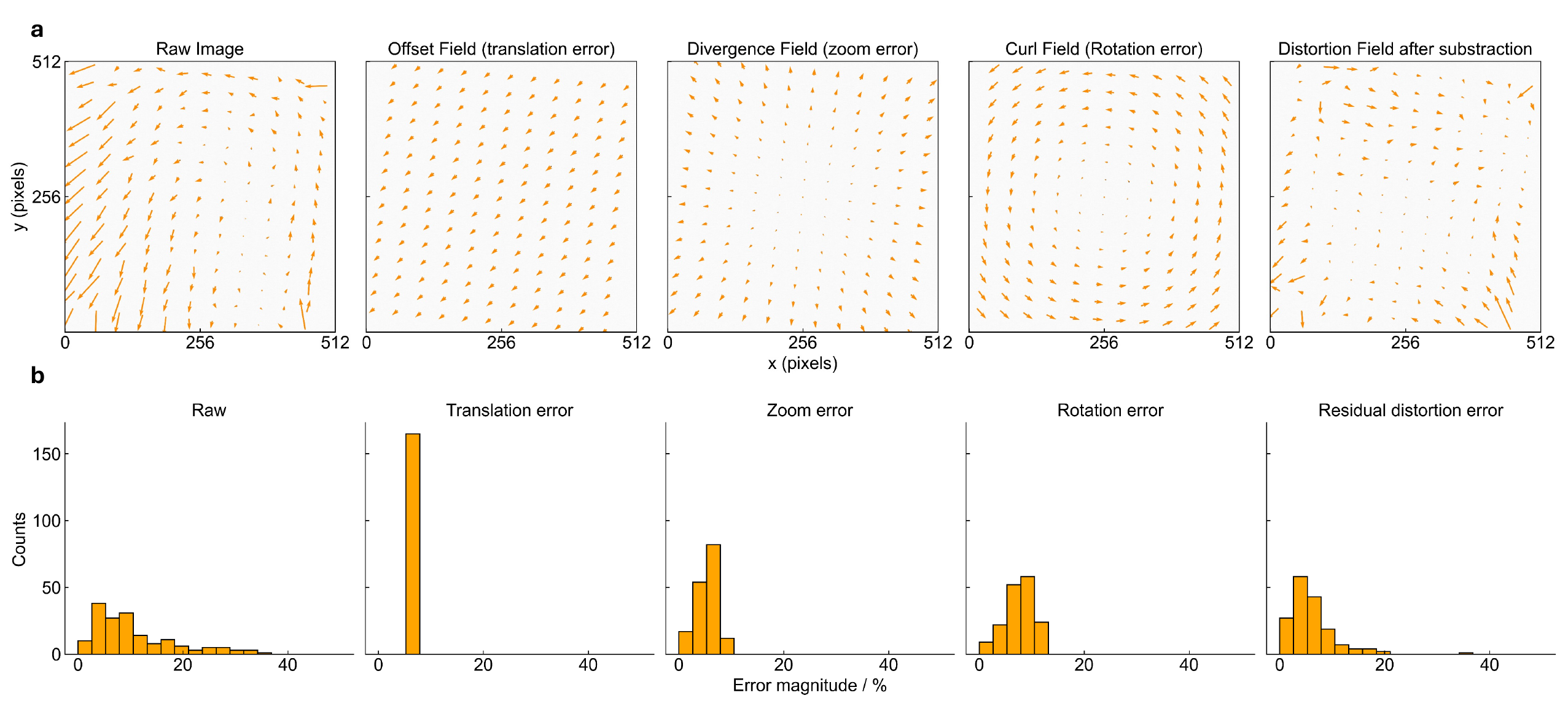}
	\caption{\textbf{Distortion Measurement.} \textbf{(a)} After deriving the discrete vector field, the alignment error caused by linear operations, including translation, rotation, and zoom, are extracted from the mean offset, curl, and divergence of the vector field, respectively. The residual error is caused by nonlinear operations, such as optical aberrations. \textbf{(b)} Histogram of magnitude of the vectors in \textbf{(a)} with a residual error smaller than 20\%.}
	\label{fig:distortion2}
\end{figure*}

Given the background-subtracted intensity value of grid points, $I(N_x,N_y)$, we first compute the "misalignment vector", $v_x$ and $v_y$, by computing the center of mass of intensity distribution taking into account the four adjacent grid points:
\begin{gather}
    I_{\text{total}}=I(N_x+1,N_y)+I(N_x-1,N_y)+I(N_x,N_y+1)+I(N_x,N_y-1)+I(N_x,N_y)\\
    v_x(N_x,N_y)=\dfrac{I(N_x+1,N_y)-I(N_x-1,N_y)}{I_{\text{total}}}\\
    v_y(N_x,N_y)=\dfrac{I(N_x,N_y+1)-I(N_x,N_y-1)}{I_{\text{total}}}
\end{gather}
In other words, we compute the center of mass based on intensity distributions around each grid point and represent the position as a 2D vector. After computing the misalignment vectors for all the grid points, we calculate the discrete vector field on the grid. Then the misalignment caused by translations can be extracted by computing the mean value (offset) of the vector field, the mismatch of the zoom ratio of the macro lens can be computed as the mean divergence of the vector field, and the rotation misalignment can be computed as the mean curl of the vector field. We view the remaining vector field as the nonlinear distortion caused by the optical system, for example optical aberrations. In our experiment the residual distortion is typically smaller than 20\% (Fig.~\ref{fig:distortion2}).

\section{Simple Model of Time Contrast Behavior}
\label{SI:contrastbehavior}
\begin{figure}[t]
	\centering
 	\includegraphics[width=\textwidth]{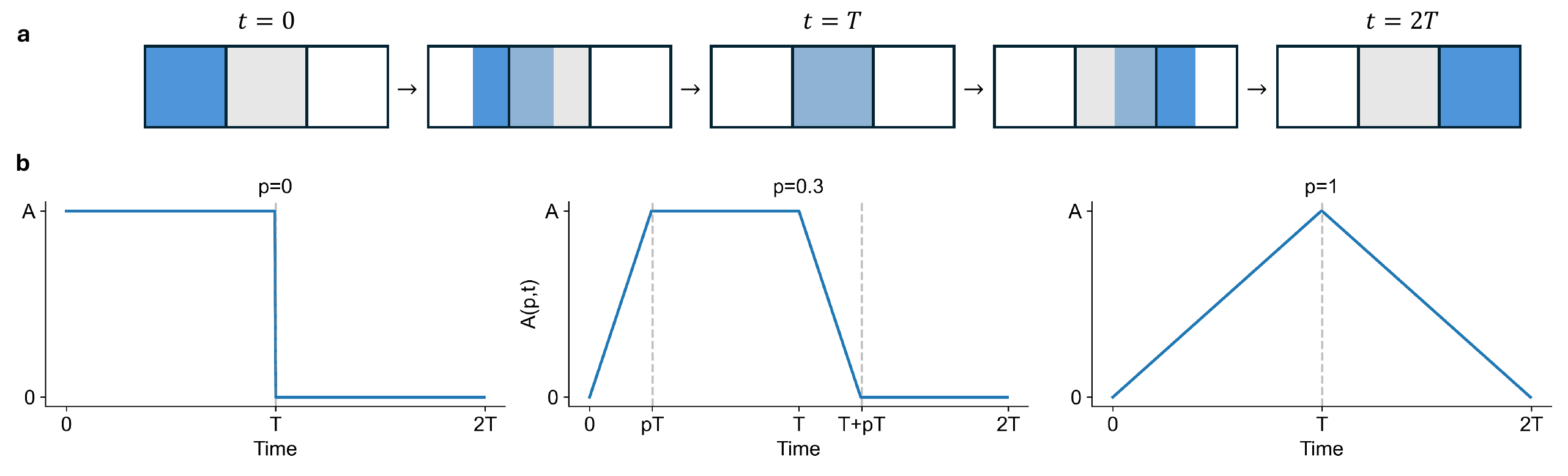}
	\caption{\textbf{Demonstration of charge transfer dynamics.} \textbf{(a)} Schematic of proposed charge shifting process. Here the charges (blue) are shifted from left to right (from time $t=0$ to $t=2T$, $T=1/f_{\text{vert}}$), and center gray region represents the only exposed single pixel. In our model we assume that the charges transfer at a constant velocity, and the charges are being exposed while being shifted so the effective exposure area (overlap between gray area and blue area) changes during the transfer. \textbf{(b)} Time plot of effective exposure areas at different $p$. Any non-zero transfer time will result in the actual integration time for each data point longer than a time bin, which will reduce the contrast on AM modulated input signal. Note that at extreme case of $p=0$, the sampling will only take the time of $1/f_{\text{vert}}$, while at case $p=1$ we will have the sampling span the time range of $2/f_{\text{vert}}$.
	}
	\label{fig:chargetransferdynamics}
\end{figure}

The charge transfer dynamics strongly impact the camera’s time contrast behavior, as it defines the effective integration time per frame. In order to understand the behavior in different operating modes, we use a simple model for the charge transfer dynamics that considers the portion of charge shifting time during the cycle, denoted as $p$, and the pixel spill-out $s$. For simplicity we assume that the effective exposure area changes linearly during the charge row-shift. The expressions of the effective area $A(p,t)$ at time $t$ during two working cycles is given as (see Fig.~\ref{fig:chargetransferdynamics}):
\begin{equation}
    A(p,t)=\left\{\begin{matrix}
 \dfrac{A}{pT}t  & 0\leq t <pT\\
 A & pT\leq t <T\\
 A-\dfrac{A}{pT}(t-T)  & T\leq t < T+pT \\
 0 & T+pT\leq t <2T
\end{matrix}\right.
\end{equation}
where $T=1/f_{\text{vert}}$ is the time period for one shifting cycle and $A$ is the area for each pixel. This shows that for a finite charge transfer time, each data point will sample a time period that spans longer than a time bin, which will in return reduce our time modulation contrast. 

To simulate the contrast at a given modulation frequency and row shifting rate, we first compute the counts $I(nT)$ for all time bins, denoted by $nT$, assuming no spill-out:
\begin{equation}
    I(nT)=\int_{nT}^{(n+2)T}A(p,t)S(t)dt
\end{equation}
Here, $S(t)$ denotes the experimentally measured square-wave like input signal. Afterwards, we calculate the spill-out-affected signal $I_s(nT)$ as
\begin{equation}
    I_s(nT)=sI\left[(n-1)T\right]+I(nT)+sI\left[(n+1)T\right]
\end{equation}
from which we extract the contrast of the simulated signal by computing the difference between maximum and minimum value of the signal, equivalent to the procedure we use for the experimental data. For a given input signal, $S(t)$, and shifting frequency, $f_{\text{vert}}=1/T$, the contrast is therefore a function of $s$ and $p$. We measure the spill-out $s$ in a separate experiment, hence only $p$ serves as fitting parameter. The fitting results are $p$=0.44 for 2~MHz normal mode, $p$=0.49 for 1.67~MHz FKM mode, and $p$=0.38 for 3.33~MHz normal mode as shown in Fig.~\ref{fig:ContrastFig}.  The good agreement of the fitting curves with the data suggests that indeed the charge shifting operation's dynamics play an important role.

\section{Scalability of the system}

In this work, we demonstrate 1.67~MHz video acquisition with a resolution of 11$\times$15. However, for many applications a larger resolution is desirable. In this section we describe the trade-off between resolution and the maximum number of frames.

\begin{figure}[h]
	\centering
 	\includegraphics[width=\textwidth]{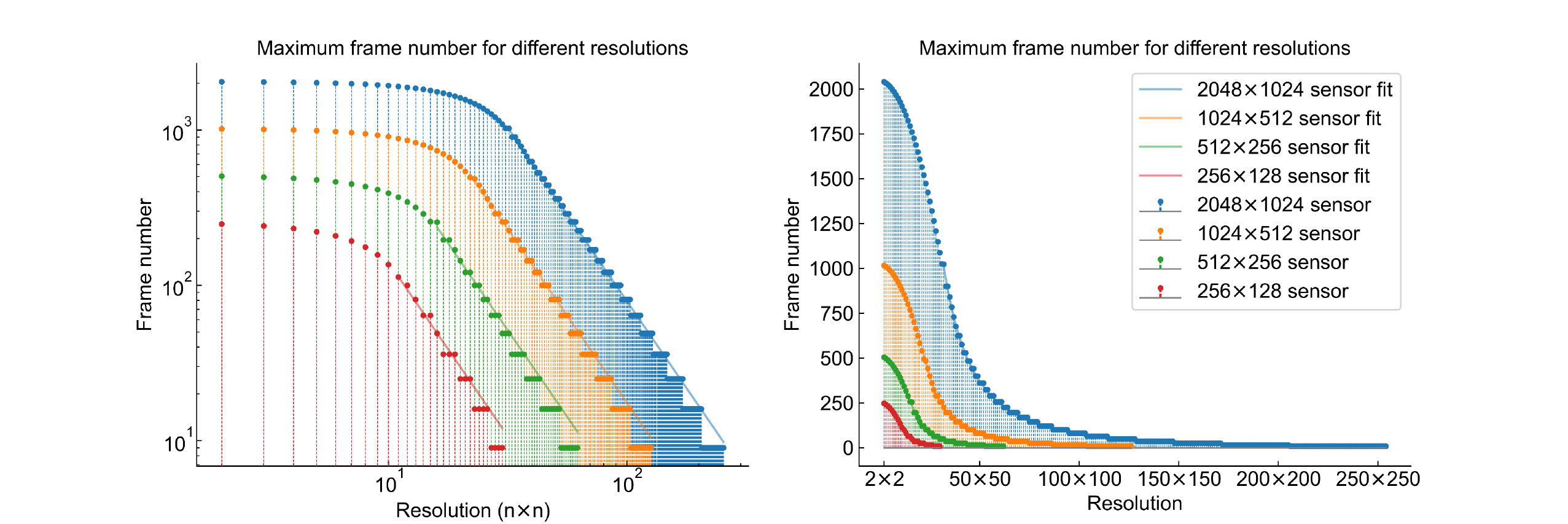}
	\caption{\textbf{Demonstration on the scalability of the system.} \textbf{(left)} and \textbf{(right)} are the same plots in log and decimal scales respectively.}
	\label{fig:scalability}
\end{figure}

We assume a frame transfer CCD, in which both the exposure area and storage area can be read out and approximate the total sensor size as $2N_s \times N_s$. We now compute the condition an $N\times N$ square grid (stemming from a square pitch lens array) must satisfy to form a single pixel grid on the sensor. We follow the symbol definition in SI~\ref{SI:Distortion} and denote $l$ as the length of the pitch and set $R=1$, $p=1$ for simplicity. Under these conditions, a grid can be constructed by specifying $N$, $N_h$, and $N_h'$, which satisfy the following equations:
\begin{equation}
	\left\{\begin{matrix}
		l\cos \theta=N_h\\
		l\sin\theta=N_h'\\
	   (N-1)l\cos\theta+(N-1)l\sin\theta=N_s-1
	   \end{matrix}\right.
\end{equation}
Notice that under square grid condition $N_h=N_v$ and $N_h'=N_v'$. Solving this equation results in:
\begin{equation}
	(N-1)(N_h+N_h')=N_s-1
\end{equation}
Given that $(N_h+N_h')$ must always be integer, and $N_s-1$ is not always divisible by $N$, we compute:
\begin{equation}
	N_h+N_h'=\lfloor \frac{N_s-1}{N-1} \rfloor
\end{equation}
and iterate through all the $N_h$s and $N_h'$s. After specifying the exact grid, we can compute the maximum frame number by calculating the distance between the bottom grid point and the nearest grid point in the same column. For $N> \left\lfloor\sqrt{N_s-1} \right\rfloor$ this numerical method indicates that the frame number will scale as $N^{-2.2}$. For the case that $N\leq\left\lfloor\sqrt{N_s-1} \right\rfloor$, we will instead simply compute the most compact grid without overlapping, i.e., $N_h'=1$, $N_h=N$, and then extract the vertical size of it. For this case, the number of frames is the difference between sensor size and the vertical size of the grid, which we compute analytically as $2N_s-N^2$. Fig.~\ref{fig:scalability} shows that for our 1024$\times$512 sensor, we are able to acquire video with $\sim$20 frames and $100\times 100$ resolution. However, at the same resolution a  2048$\times$1024 sensor allows to take $\sim100$ frames.
\label{SI:scalability}

\clearpage
\putbib
\end{bibunit}

\end{document}